\documentclass[11pt,twoside]{article}
\usepackage{asp2010}

\resetcounters

\usepackage[english]{babel}
\usepackage{graphicx}
\usepackage{bm}
\usepackage{booktabs}
\usepackage{color}
\usepackage{hyperref}
\usepackage{latexsym}
\usepackage{natbib}
\usepackage{times}
\usepackage{stmaryrd}

\newcommand{\eg}{\textit{e.g.}~}

\begin{document}
\title{High-Order Numerical-Relativity Simulations of Binary Neutron
Stars}
\author{David Radice$^1$, Luciano Rezzolla$^2$ and Filippo Galeazzi$^2$}
\affil{$^1$ TAPIR, Walter Burke Institute for Theoretical Physics, California
Institute of Technology, Pasadena, CA 91125, USA}
\affil{$^2$ Institut f\"ur Theoretische Physik, Max-von-Laue-Str. 1,
  60438 Frankfurt am Main, Germany}

\begin{abstract}
We report simulations of the inspiral and merger of binary neutron stars
performed with \texttt{WhiskyTHC}, the first of a new generation of
numerical relativity codes employing higher than second-order methods for
both the spacetime and the hydrodynamic evolution. We find that the use
of higher-order schemes improves substantially the quality of the
gravitational waveforms extracted from the simulations when compared to
those computed using traditional second-order schemes. The reduced
de-phasing and the faster convergence rate allow us to estimate the
phase evolution of the gravitational waves emitted, as well as the
magnitude of finite-resolution effects, without the need of phase- or
time-alignments or rescalings of the waves, as sometimes done in other
works. Furthermore, by using an additional unpublished simulation at very
high resolution, we confirm the robustness of our high convergence order
of $3.2$.
\end{abstract}

\section{Introduction}
The inspiral and merger of binary neutron stars (BNSs) is one of the most
promising sources of gravitational waves (GWs) for future ground-based
laser-interferometer detectors such as LIGO, Virgo or KAGRA
\citep{Sathyaprakash:2009xs}. Because they can travel almost unscattered
through matter, GWs carry valuable information from the deep core of the
neutron stars (NSs) concerning the equation of state (EOS) of matter at
supra-nuclear densities. Unfortunately, they are also extremely hard to
detect, so that their identification and analysis requires the
availability of analytical or semi-analytical GW templates. In turn, the
validation and tuning of these models must be done by matching them with
the predictions of fully non-linear numerical relativity (NR)
calculations, which represent the only means to describe accurately the
late inspiral of BNS \citep{Baiotti:2010, Baiotti2011, Bernuzzi2012,
Hotokezaka2013b, Bernuzzi2014, Bernuzzi2015}.

While very high-quality NR waveforms of binary black hole mergers are
available, \eg \citep{Aylott:2009ya, Mroue2013, Hinder2013}, BNS
simulations have been plagued by low convergence order and relatively
large phase uncertainties ($\delta\phi/\phi \sim 1 \%$)
\citep{Baiotti:2009gk,Bernuzzi2011}.  Furthermore, since NSs have smaller
masses, the merger part of the waveform is out of the frequency band for
the next generation GW detectors, so that EOS-related effects will have
to be most probably extracted from the inspiral signal using a large
number of events.  In particular, EOS-induced effects will be encoded in
the phase evolution of the GW signal during the inspiral
\citep{Damour:2012}. As a result, the measure of EOS-induced effects
requires very accurate general-relativistic predictions of the inspiral
signal. Even though accurate waveforms can be calculated by second-order
codes at very high computational costs \citep{Baiotti:2010, Baiotti2011,
  Bernuzzi2012, Hotokezaka2013b, Bernuzzi2014, Bernuzzi2015}, their
analysis is complicated by the low convergence order of the methods
employed. In particular, the analysis often requires the use of a time
rescaling or alignment of the GWs from different resolutions
\citep{Baiotti2011, Hotokezaka2013b}, which is hard to justify
mathematically.

Here we show that, by using high-order numerical methods, it is indeed
possible to obtain waveforms for the late-inspiral of a BNS system with a
quality that is almost comparable with the one obtained for binary black
holes, with clean, higher than second-order convergence in both the phase
and the amplitude.  In particular we highlight and extend to higher
resolution some of the results we reported in \citep{Radice2013b}.

\section{Numerical methods}

The results presented here have been obtained with our new high-order,
high-resolution shock-capturing (HRSC), finite-differencing code:
\texttt{WhiskyTHC} \citep{Radice2013b, Radice2013c}, which represents the
extension to general relativity of the special-relativistic \texttt{THC}
code \citep{Radice2012a}. \texttt{WhiskyTHC} solves the equations of
general-relativistic hydrodynamics in conservation form
\citep{Banyuls97,Rezzolla_book:2013} using a finite-difference scheme
that employs flux reconstruction in local-characteristic variables using
the MP5 scheme, formally fifth-order in space \citep{suresh_1997_amp}
[see \citet{Radice2012a, Radice2013c} for details].

The spacetime evolution makes use of the BSSNOK formulation of the
Einstein equations \citep{Nakamura87, Shibata95, Baumgarte99} and it is
performed using fourth-order accurate finite-difference scheme provided
by the \texttt{Mclachlan} and is part of the \texttt{Einstein Toolkit}
\citep{Loffler:2011ay, Brown:2008sb, Schnetter-etal-03b}. To ensure the
non-linear stability of the scheme we add a fifth-order Kreiss-Oliger
type artificial dissipation to the spacetime variables only. Finally, the
coupling between the hydrodynamic and the spacetime solvers is done using
the method of lines and a fourth-order Runge-Kutta time integrator. The
resulting scheme is formally fourth-order in space and time, except at
the boundaries between different refinement levels, where our method is
only second-order in time. This should have only marginal effects on our
results given that our finest grid covers both NSs.

\section{Binary setup}
\label{sec:setup}
\begin{table}
\caption{\label{table:models} Summary of the considered BNS model. We
  report the total baryonic mass $M_b$, the ADM mass $M$, the initial
  separation $r$, the initial orbital frequency $f_{\mathrm{orb}}$, the
  gravitational mass of each star at infinite separation,
  $M_\infty$, the compactness, $\mathcal{C} = M_\infty/R_\infty$, where
  $R_\infty$ is the areal radius of the star when isolated and the tidal
  Love number, $\kappa_{2}$, \eg \citet{Hinderer09}.}
\vspace{1em}
\centering
\begin{tabular}{ccccccc}
  \tableline
  \noalign{\smallskip}
    $M_b\ [M_\odot]$ &
    $M\ [M_\odot]$ &
    $r\ [\mathrm{km}]$ &
    $f_{\mathrm{orb}}\ [\mathrm{Hz}]$ &
    $M_\infty\ [M_\odot]$ &
    $\mathcal{C}$ & 
    $\kappa_2$ \\
  \tableline
  \noalign{\smallskip}
  $3.8017$ & $3.45366$ & $60$ & $208.431$ & $1.7428$ & $0.18002$ & $0.05$ \\
  \tableline
  \noalign{\smallskip}
\end{tabular}
\end{table}

The initial data is computed in the conformally flat approximation using
the \texttt{\textsc{Lorene}} pseudo-spectral code \citep{Gourgoulhon01}
and describes two irrotational, equal-mass NSs in quasi-circular
orbit. Its main properties are summarized in Table \ref{table:models},
and we note that it is computed using a polytropic EOS ($p=K\rho^\Gamma$)
with $K = 123.56$ (in units where $G = M_\odot = c = 1$) and $\Gamma=2$,
while the evolution is performed using the ideal-gas EOS
($p=(\Gamma-1)\rho\epsilon$) with the same $\Gamma$
\citep{Rezzolla_book:2013}.

The runs are performed on a grid covering $0 < x,z \lesssim
750\ \mathrm{km}$, $-750\ \mathrm{km} \lesssim y \lesssim
750\ \mathrm{km}$, where we assume reflection symmetry across the $(x,y)$
plane and $\pi$ symmetry across the $(y,z)$ plane. The grid employs six
\emph{fixed} refinement levels, with the finest one covering both stars.
We consider four different resolutions, labelled as $L$, $M$, $H$ and
$V\!H$, having, in the finest refinement level, a grid spacing of
$h\simeq 370, 295, 215$ and $147$ meters, respectively. The results of
simulations $L$, $M$, $H$ were already presented in \citet{Radice2013b,
  Radice2013c}, while those of the simulation $V\!H$ are presented here
for the first time.

\section{Binary dynamics}
\begin{figure}
  \begin{center}
    \includegraphics[width=\hsize]{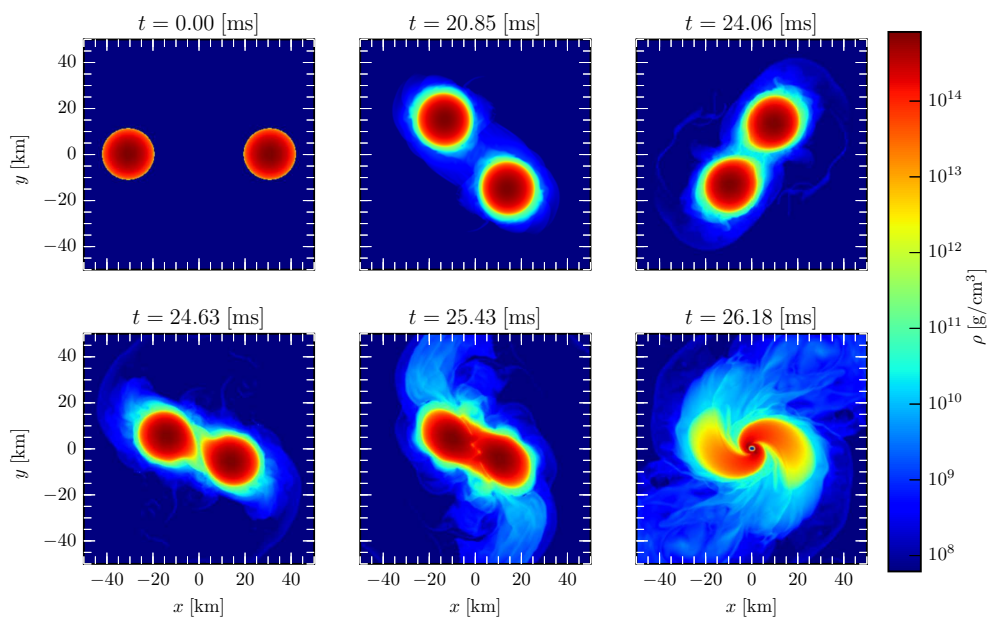}
    \caption{Rest-mass density on the equatorial plane for the medium
      resolution run $M$. The panels show the initial data (upper right),
      the late stages of the inspiral (top middle and left), the
      approximate time of contact (bottom left), merger (bottom center)
      and black-hole formation (bottom right).}
    \label{fig:rho2d}
  \end{center}
\end{figure}
The dynamics of the binary is summarized in Figure \ref{fig:rho2d}, where
we show the density on the equatorial plane for the $M$ run (the others
are qualitatively very similar). The initial distance between the two NS
centers is $60\ \mathrm{km}$. They complete about $7$ orbits, while
inspiraling because of the loss of orbital angular momentum to GWs,
before entering into contact at time $t\simeq 24.6\ \mathrm{ms}$ (from
the beginning of the simulation). Finally, they quickly merge and a black
hole is formed soon after.

\section{Gravitational waves}
\begin{figure}
  \begin{center}
    \includegraphics[width=0.49\hsize]{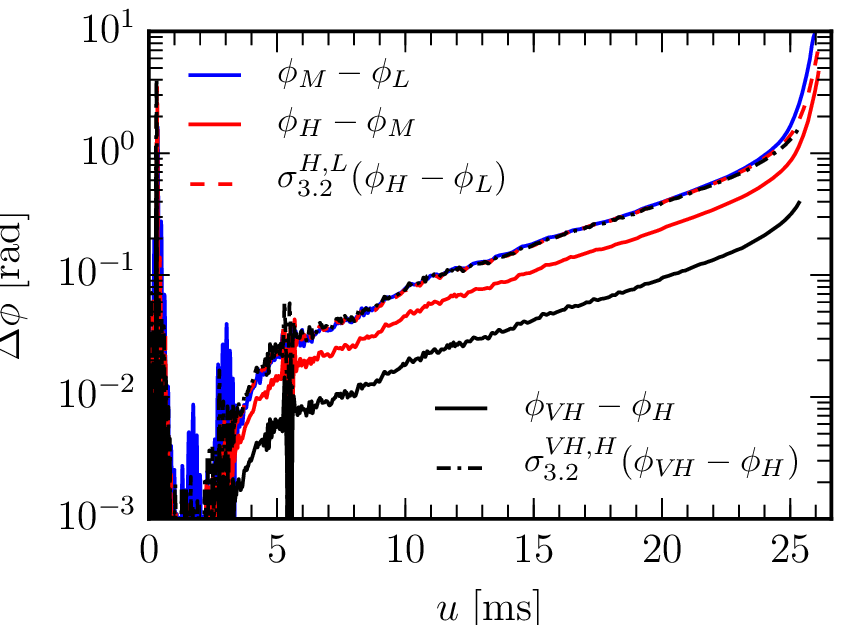}
    \includegraphics[width=0.49\hsize]{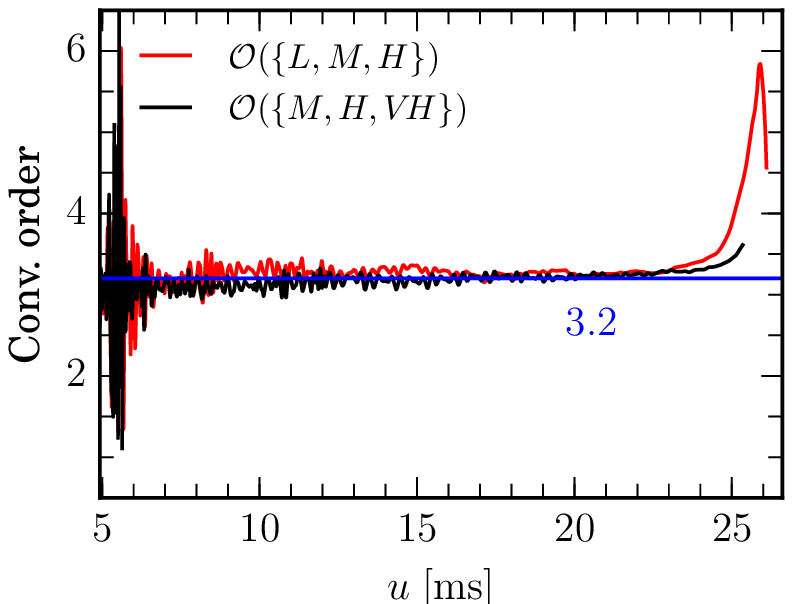}
    \caption{Accumulate de-phasing (left panel) and estimated order of
      convergence (right panel) for the $\ell = 2, m = 2$ mode of the
      Weyl scalar $\Psi_4$ as extracted at $r = 450\ M_\odot$. The
      de-phasing between high ($H$) and medium ($M$) and very high
      ($V\!H$) and high are also rescaled assuming an order of
      convergence of $3.2$. The instantaneous order of convergence is
      estimated separately from the first three resolutions
      $\mathcal{O}(\{L,M,H\})$ and from the last three
      $\mathcal{O}(\{M,H,V\!H\})$.}
  \label{fig:convergence}
  \end{center}
\end{figure}

To quantify the accuracy of our code for GW astronomy, we consider the
phase evolution of the dominant $\ell = 2, m = 2$ component of the
curvature GW, the (complex) Weyl scalar $\Psi_4$, as extracted at $r
\simeq 665\ \mathrm{km}$. We compute the GW phase $\phi$ after
decomposing the curvature as $\Psi_4 = A
\mathrm{e}^{-\mathrm{i}\phi}$. Figure \ref{fig:convergence} shows an
analysis of the residual of the phase between the different resolutions
as a function of the retarded time $u$. On the left panel, we show both
the absolute de-phasing between successive resolutions and the residuals,
between the $H$ and $M$ and the $V\!H$ and $H$ resolutions, scaled
assuming a convergence order of $3.2$.  On the right panel, we plot the
instantaneous convergence order as measured from three out of the four
resolutions (separately the first three $L, M, H$ and the last three $M,
H, V\!H$).

Figure \ref{fig:convergence} demonstrates that higher than second-order
convergence can be achieved for numerical relativity simulations of BNS,
without the need to perform any artificial manipulation of the waveforms.
We find an order of convergence $\sim 3.2$ (also confirmed by the very
good overlap between the rescaled de-phasing), which is somewhat smaller
than the formal order of four of our scheme. However, this is to be
expected because HRSC methods typically reach their nominal convergence
order only at very high resolutions \citep{Shu97, Radice2012a}; see also
\cite{Zlochower2012} for a discussion of other possible sources of
errors. Finally, as also observed with other codes \citep{Bernuzzi2011},
our solution shows a loss of convergence (with apparent
super-convergence) after $u \gtrsim 24.6\ \mathrm{ms}$. This is roughly
the time when the two NSs enter into contact (see Figure
\ref{fig:rho2d}). At this time the de-phasing between the $H$ and $V\!H$
resolution is $\simeq 0.26\ \mathrm{rad}$.

\begin{figure}
  \begin{center}
    \includegraphics[width=0.49\hsize]{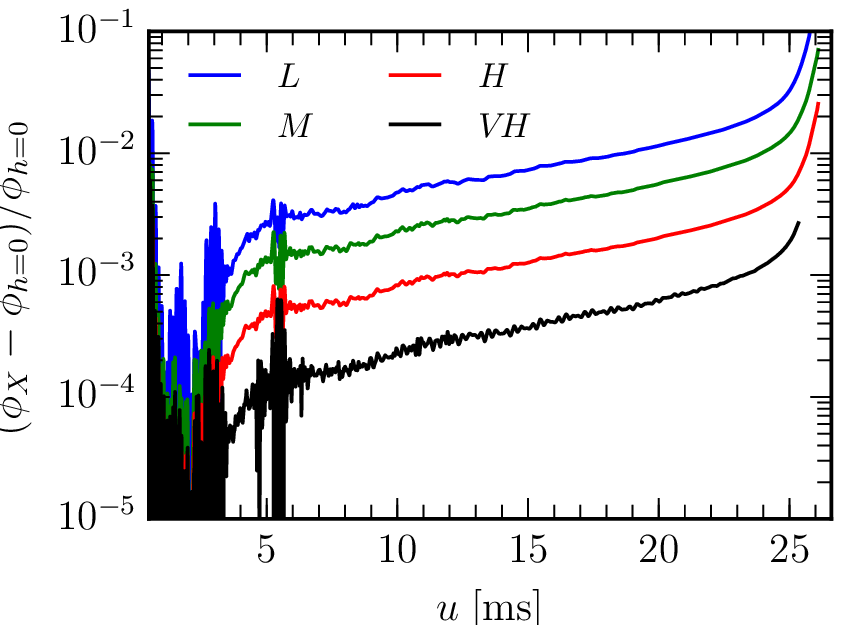}
    \includegraphics[width=0.49\hsize]{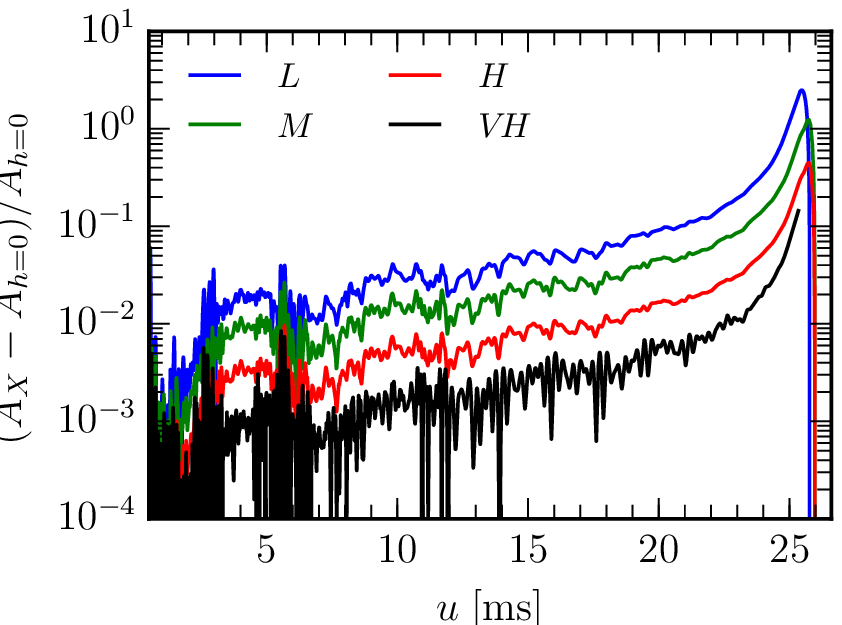}
  \caption{Relative phase (left panel) and amplitude (right panel)
  differences for the $\ell = 2, m = 2$ mode of the Weyl scalar $\Psi_4$
  between the Richardson extrapolated phase and each of the resolutions.
  The extrapolated phase is computed assuming a convergence order of
  $3.2$ and using the first three resolutions ($L$, $M$ and $H$).}
  \label{fig:error}
  \end{center}
\end{figure}

As a measure of the error with respect to the exact solution we use the
\emph{first three} resolutions ($L$, $M$ and $H$), Richardson extrapolate
the GW to infinite resolution assuming a convergence order of $3.2$ and
compute the phase and amplitude differences of each run with respect to
the extrapolated data.  The results are shown in Figure \ref{fig:error}.
Obviously the data for $u > 24.6\ \mathrm{ms}$ (which is about $13.5$ GW
cycles) has to be taken with a grain of salt, given that convergence is
lost after contact. Before that, we find that the $H$ run has a
de-phasing $\simeq 0.35\ \mathrm{rad}$ ($\simeq 0.4 \%$) at time $u =
24.6\ \mathrm{ms}$, while the highest resolution $V\!H$ run has a
de-phasing as small as $\simeq 0.13\ \mathrm{rad}$. This corresponds to a
phase error of less than $0.15 \%$, which would be challenging to achieve
with standard second-order numerical-relativity codes. The relative
errors of the amplitude are somewhat larger, but also on a few percent level for
the $H$ and $V\!H$ resolutions, which is more than adequate giving that
the amplitude is not as critical as the phase for GW astronomy. Overall,
this shows that the Richardson extrapolated waveform is consistent with
the $V\!H$ data (which lies between the extrapolated waveform and the $H$
data).

\section{Conclusions}
We presented a set of four, high-resolution, high-order, simulations of
BNS inspiral in general relativity, the highest of which has not been
published before. Our analysis focused on the accuracy of the
gravitational wave signal extracted from the simulations and, in
particular, on its phase evolution, which represents a crucial quantity
for both detection and parameters extraction.

We showed that, with the use of higher-order methods, it is possible to
achieve clean convergence and small phase errors, without the need to
perform any alignment or rescaling of the GWs. The
Richardson-extrapolated waveforms appear robust as the resolution
increases, giving support to their accuracy. This opens the possibility
of obtaining high-quality waveforms with reliable error estimates that
could be used to verify and calibrate analytical and phenomenological
phasing models to be used in GW astronomy.

\acknowledgements 

We thank W.\ Kastaun for providing the primitive recovery routine and I.\
Hawke, S.\ Bernuzzi, D.\ Alic, R.\ Haas and K.\ Takami for useful
discussions.  Partial support comes from the Sherman Fairchild
Foundation, the DFG grant SFB/Transregio 7, by ``NewCompStar'', COST
Action MP1304, and by the Helmholtz International Center for FAIR.  The
calculations were performed on SuperMUC at the LRZ, on Datura at the AEI,
and on LOEWE in Frankfurt. 

\bibliography{Radice}

\end{document}